\newcommand{\beq}{\begin{quote}}
\newcommand{\enq}{\end{quote}}
\newcommand{\be}{\begin{eqnarray}}
\newcommand{\en}{\end{eqnarray}}
\newcommand{\eps}{\epsilon}
\newcommand{\del}{\delta}

\newcommand{\Om}{\Omega}
\newcommand{\De}{\Delta}

\newcommand{\la}{\lambda}
\documentclass[preprint,showpacs,preprintnumbers,amsmath,amssymb]{revtex4}

\usepackage{graphicx}
\usepackage{dcolumn}
\usepackage{bm}

\begin{document}
\title{ 
The evolution of radiation towards thermal equilibrium:
A soluble model which illustrates the foundations of Statistical Mechanics.
} 
\date{May 9, 2003}
\author{Michael Nauenberg\\
Department of Physics\\
University of California, Santa Cruz, CA 95064 
}
\email{michael@mike.ucsc.edu}

\begin{abstract}
In 1916 Einstein introduced the first rules for a quantum
theory of electromagnetic radiation, and he applied them  to
a model of matter in thermal equilibrium with radiation    
to derive Planck's black-body formula. 
Einstein's treatment is extended here  to 
time-dependent  stochastic variables, which leads to 
a master equation for the probability distribution
that describes the irreversible approach 
of Einstein's model  towards thermal equilibrium, and
elucidates   aspects of the foundation of statistical mechanics.
An analytic solution of this equation is obtained  
in the Fokker-Planck approximation 
which is in excellent agreement with numerical results.  
At equilibrium, it is shown that the  probability distribution  
is proportional to the total number of microstates for a given configuration, 
in accordance with  Boltzmann's fundamental  postulate of 
equal { \it a priori} probabilities for these states.  
While the counting of these  configurations depends on  particle statistics-
Boltzmann, Bose-Einstein, or  Fermi-Dirac - the corresponding probability
is determined here by the {\it dynamics} which are  embodied in the form of 
Einstein's  quantum transition probabilities 
for the  emission and absorption of radiation.
In a special limit, it is shown that  the  photons 
in Einstein's  model can act as a thermal bath for the evolution of 
the  atoms towards the canonical  equilibrium  
distribution of Gibbs. In this limit, the present model is 
mathematically equivalent to an extended version of the  Ehrenfests' ``dog-flea''
model, which has been  discussed recently in this 
journal by Ambegaokar and Clerk. 

\end{abstract}
\maketitle

\subsection*{Introduction}

In a seminal paper written in 1877,  Boltzmann\cite{boltz1}  
formulated the basic principle of statistical mechanics: 
in equilibrium  all of the microstates of an isolated  system  that have the  same total  energy 
occur with equal  {\it a priori} probability. 
As a consequence of this principle, the probability for a given macroscopic configuration 
is proportional to the accessible  number of microscopic  states 
which comprises it, and Boltzmann demonstrated  that the thermodynamic
entropy is proportional to the logarithm of the maximum  value
of this number of configurations.  Boltzmann's  principle is  often  
regarded as a {\it necessary}  postulate 
from which  other concepts of statistical mechanics  
and its relationship to  thermodynamics can be deduced \cite{tolman1}.
In 1900 Planck made a fundamental  application of this postulate 
in deriving  his famous  black-body radiation formula \cite{planck1}.
Planck considered  the  source of the  radiation  to be
Hertzian electromagnetic oscillators of 
frequency $\nu$, but  he introduced  the radical assumption 
that the energies of these oscillators were  quantized in units of  
an energy element  $\epsilon=h\nu$, where $h$ is  Planck's constant.
This assumption yields a finite and  countable number of microscopic 
configurations of fixed energy for which  Boltzmann's statistical postulate
could be readily applied \cite{planck2}.
Then  in 1916  Einstein developed a different approach in
describing the condition for the thermal equilibrium of radiation and matter which
lead him also to a derivation of  Planck's black-body radiation formula \cite{einst0}.
Following  Bohr's quantum rules  for the atom, Einstein  
proposed a quantum theory for the emission and absorption of  radiation,
and he then  combined it with  a detailed balance argument
to obtain Planck's  equilibrium distribution for thermal radiation \cite {einst1}.

But why should these two completely  different derivations,
which  have in common only Planck's  quantum  hypothesis  that the 
energy of the source of radiation is {\it discrete}, and otherwise are based on 
different physical and mathematical assumptions, 
lead to the same result? 
Surprisingly, it appears that this question was not addressed in the past, 
and presently it is generally ignored in textbooks on statistical 
mechanics \cite{kittel}.
In this paper we propose to answer this question by  considering 
a stochastic treatment of Einstein's model. 
It will be shown that the equilibrium probability  distribution
obtained this way  is the same as the  result obtained by applying  
Boltzmann's postulate  of equal a priori probabilities for
the microstates of a system composed of atoms and photons 
with fixed total energy.
While the counting of these  configurations depends on  particle statistics, 
which can be Boltzmann, Bose-Einstein, or Fermi-Dirac, the associated  probability 
distribution is determined in our treatment 
by the {\it dynamics} which are  embodied in the 
Einstein's  transition probabilities for emission and absorption of radiation.
We obtain also a time dependent  master  equation \cite{vankampen} that 
describes the irreversible approach of  radiation and matter  towards thermal equilibrium.
In the Fokker-Planck approximation  we obtain an  analytic solution  
in excellent agreement with numerical results some of which are  presented here. 

In section  we present a historical account of Einstein's model
for the thermal equilibrium of radiation with matter, and we describe the
detailed  balance argument which he used  to derive Planck's formula
for black-body radiation. In section II
we extend this treatment  by assuming that the  variables   in
Einstein's model are stochastic variables, and we derive a 
probability function that describes the stationary  equilibrium
of this model.  We show that this 
probability function is proportional to  
the number of configurations of  the atoms and photons
at fixed total energy, in accordance with Boltzmann's fundamental postulate
of statistical mechanics
and the symmetries based on quantum statistics.
Thus, the counting associated with  Bose-Einstein statistics 
which gives the  probability 
distribution for photons in statistical mechanics 
\cite{bose1} \cite{einst2}, 
is obtained  here from the quantum 
dynamics  embodied in  Einstein's  transition probabilies for the
absorption and emission of photon by atoms.
Likewise, we obtain the corresponding results associated with 
Boltzmann's statistics for atoms that are treated as distinguishable particles,
or with Fermi-Dirac statistics for atoms that satisfy Pauli's exclusion principle. 
In section III we obtain  the most probable  configuration 
by evaluating  the maximum  of the equilibrium
probability function. This also  corresponds  to  the 
maximum  value  of Boltzmann's  entropy function, which is 
the justification in statistical mechanics  
of the second law of thermodynamics.
In section IV we consider a {\it master equation} \cite{vankampen} for the 
probability distribution  that describes
the time evolution of Einstein's model when  it is initially in an arbitrary  non-equilibrium
state, and we show that it always approaches a unique
function  which is the equilibrium probability
obtained in section II.  Defining  an appropriate  entropy function 
that increases monotonically in time ( a proof is given in Appendix A), we show that  
its  maximum value  determines the stationary or
equilibrium probability. The time evolution and the  approach to
equilibrium of the probability function is also illustrated by 
numerical solutions of this master equation.
In section V we approximate the master equation by
a Fokker-Planck equation, and we  obtain an  analytic solution 
in the form of a Gaussian function with a time-dependent mean value
and a width parameter which is found to be in excellent agreement with our numerical solutions.
Finally we  consider  a special limiting case where the photons act as a thermal bath
for the atoms, which corresponds to an extension of a statistical model for
the approach to thermal  equilibrium by Paul and Tatiana Ehrenfest 
as extended recently by Ambegaokar and Clerk \cite{ambegaokar}
A summary and some conclusions are presented in section VI.

\subsection *{I. Historical background: Einstein's  quantum theory of radiation  and its 
application  to  atoms  and photons in thermal equilibrium}

In 1916 Einstein introduced a  quantum theory for the emission and
absorption of electromagnetic radiation by atoms which  anticipated
the modern theory of quantum electrodynamics, and  he  applied it  to a derivation of 
Planck's black-body distribution for thermal  radiation \cite{einst0}. 
His  model for matter  in thermal equilibrium with radiation 
consisted  of atoms or molecules  with discrete energy levels, 
which exchange energy with electromagnetic radiation contained  inside   
a cavity with reflecting walls.
For simplicity we assume here that these atoms  have only
two levels with an energy difference $\eps$, and that the radiation 
has frequencies in the range $\nu, \nu+d\nu$. 
Instead of applying Maxwell's  classical  electromagnetic theory,  
Einstein  assumed that the  interaction of radiation  with matter
gives  rise  to quantum transitions between the energy levels 
of these atoms which are associated with  the absorption and 
emission of radiation. He introduced 
transition probabilities  per unit time for the stimulated  absorption  
and emission of radiation  by an atom which
are proportional 
to the electromagnetic energy density $\rho_{\nu}$ in the cavity.
in  analogy with the  exchange of energy
between an electromagnetic  oscillator and  radiation in classical
theory.
In addition, he introduced a probability for the {\it spontaneous} 
emission of radiation from  an excited state of the atom
independent of the radiation  in  the cavity, to take into
account the classical radiation of a charged  oscillator. 
Hence  Einstein's absorption probability $p_a$  and emission probability 
$p_e$ have the form

\begin{equation}
\label{pa1}
p_a=B_a \rho_{\nu}
\end{equation}
and
\begin{equation}
\label{pe1}
p_e=A+B_e \rho_{\nu},
\end{equation}
where $B_a$ and $B_e$  are  undetermined   coefficients 
for absorption and stimulated emission respectively, and $A$ is 
the coefficient for spontaneous emission  \cite{einst1}. 

Einstein' derivation of Planck's black-body formula proceeded as follows.
He argued that in thermal equilibrium the {\it average}  number $n_g$ of atoms in the ground state 
times the probability for absorbing  radiation  
per unit time must be equal to the  average  number $n_e$ of atoms in the excited state
times the probability for emitting radiation per unit time, which leads to the 
{\it detailed balance}  equation
\be
\label{einstein1}
p_a n_g=p_e n_e.
\en
Next, he assumed that the atoms are in  thermal equilibrium at 
a temperature $T$, and he invoked the canonical Gibbs distribution
to determine the ratio
\be
\label{statmech}
\frac{n_e}{n_g}=e^{-\eps/kT}.
\en
Substituting this relation into Eq. \ref{einstein1} and using
Eqs. \ref{pa1} and \ref{pe1}, he obtain the following relation for the
equilibrium thermal radiation (black-body)  energy density  $\rho_{\nu}$  
as a function of the temperature $T$ in the cavity :

\be
\label{einstein2}
\rho_{\nu}=\frac{A}{(B_a e^{\eps/kT}-B_e)}
\en
By 1900 it was known experimentally that for large temperatures and 
long wavelengths (small frequencies)
the  black-body radiation energy spectrum depends linearly 
on the temperature. 
At about the same time, by applying the equipartition
theorem Rayleigh  also pointed out  \cite{rayleigh} that 
\be
\label{rayleigh1}
\rho_{\nu}= \gamma \nu^2 kT,
\en
where the constant  $\gamma=8\pi/c^3$ was first calculated correctly by Jeans five
years later\cite {jeans}.
For $T>>\eps/k$, this limit is  obtained from  Eq. \ref{einstein2} 
provided  that $B_a=B_e= B$, which yields the relation
\be
\label{einstein3}
\rho_{\nu}=\frac{A/B}{(e^{\eps/kT}-1)}.
\en
At this point Einstein  appealed to Wien's remarkable theoretical result that $\rho_{\nu}$
has a {\it scaling}  dependence  on the variables  $\nu$ and $T$
(Wien's displacement law) of the form
\be
\label{wien1}
\rho_{\nu}=\nu^3 f(\nu/T),
\en
where $f$  was  an  undetermined  function.
Comparing Wien's result with  Eq.\ref{einstein3},  Einstein 
deduced the frequency dependences of his two  
undetermined parameters
\be
\frac{A}{B}=\alpha \nu^3
\en
and 
\be
\label{bohr1}
\eps=h\nu,
\en
where $\alpha$ and $h$ are universal constants, and obtained
Planck's black-body formula. r

In this brilliant {\it  tour de force} in which the
probabilistic foundation of quantum theory was fist enunciated, Einstein not only derived  
Planck's formula, but through an independent route he also obtained 
the fundamental quantum relation, Eq. \ref{bohr1}, between the frequency $\nu$
of the emitted and absorbed radiation and the energy difference $\eps$  
between  two atomic levels - a relation which had been  introduced earlier  by 
Planck and thirteen years  later by Bohr. The constant $h$ is of course Planck's famous constant,
while  the constant $\alpha = 8\pi h/c^3$ had also been obtained  by Planck.
This  constant can also be obtained from the  constant $\gamma$  
in the Rayleigh-Jeans limit, Eq. \ref{rayleigh1}, which implies
that $\alpha = \gamma h$. This result was well known to Einstein, 
who had obtained  this limit from a relation of Planck 
which he described in his famous 1905 paper
on a ``heuristic'' view of light \cite{einst4}, 
but now he commented only  
that to compute the numerical value of the constant $\alpha$, one would have 
to have an exact theory of electrodynamics and mechanical processes'' \cite{einst1}.

In his original  model Einstein assumed that the atoms  could  move freely, and
he also demonstrated  that  the momentum transfer by the absorbed 
and emitted radiation was given
by $h\nu/c$ along a direction ``determined  by
`chance' according to the present state of the theory''.
In this manner  Einstein abstracted the fundamental  concept of  
{\it photons} of different frequencies  as the quantum states 
of electromagnetic radiation ``entirely from
statistical mechanics considerations'' \cite{pais1}.  
It also  follows from these considerations  that the number density $n_{\nu}$  of photons
in a radiation field of frequency $\nu$ and energy density $\rho_{\nu}$
is given by $n_{\nu}=\rho_{\nu}/h\nu$, although this relation was not stated explicitly 
in his paper.
Einstein's remarkable ``quantum  hypothesis on the radiation exchange of energy ''  
\cite{einst1} was  confirmed when the
quantum theory of electromagnetic radiation was developed,
giving  an explicit expression for the coefficient of  $B$,  
and showing  that the  ratio  $(A/B)(1/h\nu)=8\pi \nu^2/c^3$ corresponds
to the number of momentum states of the photons per unit frequency interval, 
as had been conjectured by Bose \cite{bose1}.

Since the quantum theory is invariant under time reversal, it may seem  surprising
that Einstein transition probabilities for the absorption and emission 
of radiation, Eqs.\ref{pa1} and \ref{pe1}, are  different, due
to the  extra contribution for spontaneous emission. But 
these probabilities are averages over the direction of momentum
of the transition probabilities for microstates with  photons of fixed momentum.
In the quantum  dipole approximation,
the probability of emission of a photon with momentum $\vec p$ is proportional
to $n(\vec p)+1$, where $n(\vec p)$ is the initial number of photons,
while the probability for absorption with the same initial number of photons 
is proportional to $n(\vec p)$.
When integrated over  momenta,  this  yields Einstein's transition probabilities.
For transitions between microstates, however, the proper comparison should be made by 
considering the absorption probability  for $n(\vec p)+1$ photons  in the
initial state, in which case  these transition probabilities are the same.

\subsection*{II.  Extension of Einstein's detailed balance argument to stochastic variables }

Even in thermal equilibrium, transitions associated with the
absorption and emission of photons are  occurring  continuously, 
and therefore the  number of atoms in  the ground 
state $n_g$  and in the excited state, $n_e$, as 
well as the number of photons $n_p$ in the radiation field 
must be consider to be  stochastic variables which fluctuate in time. 
Hence  the  variables  in Einstein's  
detailed balance equation, Eq. \ref{einstein1}, correspond to some  ``average''
value of these quantities. In order to  describe Einstein's  model
in further detail, we introduce a  probability function    $P_{eq}$  
for the values of these stochastic variables.
In  accordance with the  constraints of a fixed number 
of atoms $n=n_g+n_e$ and a fixed total energy $E=n_p h\nu +n_e \eps$,
we can express $P_{eq}$ as a function of a single integer $k$, where
$n_g=k$, $n_e=n-k$,  $n_p=n_q-n+k$, 
and  $n_q=E/h\nu$. For  $n \leq n_q$, the range of $k$ is   
$0\leq k  \leq n$, while  for $0 \leq n_q <n$, the corresponding
range  of $k$ is $n - n_q \leq k \leq n $. 
Then the probability that there are $k$ atoms in the ground state, 
and that  one of these 
absorbs a photon during the interval of time $\del t$ 
is given by 

\be
\label{Pa1}
Q_a(k)=k p_a(k)P_{eq}(k)\del t,
\en
while  the corresponding probability that one of the  $n-k$  atoms  in the
excited state  emits a photon is given by 
\be
\label{Pe1}
Q_e(k)=(n-k)p_e(k)P_{eq}(k)\del t.
\en
Since $n_p=\rho_{\nu} V d\nu/h\nu$, where $V$ is the volume of the
cavity, the basic transition probabilities for emission and absorption 
of photons, Eqs. \ref{pa1} and \ref{pe1}, can also be expressed in terms of the number of
photons $n_p$ in the cavity by
\be
\label{pa11}
p_a(k)=B'_a n_p
\en
and
\be
\label{pe11}
p_e(k)=B'_e(g+ n_p)
\en
where $B'_a=B_a h\nu/V d\nu$, $B'_e=B_e h\nu/V d\nu$, and $g=(8\pi/c^3) V \nu^2 d\nu$ 
is the number of photon momentum states  of frequency $\nu$  
in an  interval $d\nu$ inside  a cavity of volume $V$.

The condition for thermal  equilibrium  
requires  that  a configuration  of the system having 
$n_e=n-k$ atoms in the excited state have the same probability $Q_e(k)$ 
to emit a photon during any  time interval $\del t$  as the
probability $Q_a(k+1)$ that  a  configuration  with $n_g= k+1$ atoms in the
ground state absorb a photon during this time interval. This requirement implies  
the  extended  detailed balance relation 
\be
\label{balance1}
Q_a(k+1)=Q_e(k)
\en
It can be shown that 
if one neglects the  correlations between the number of atoms 
and the number of photons in a given state, 
Einstein's detailed  balance relation, Eq. \ref{einstein1},
can be  recovered  by summing both sides of Eq. \ref{balance1} over $k$, 
where the quantities $n_g,n_e$ and $n_g$ now refer
to averages over the distribution $P_{eq}(k)$. 

We now proceed  to solve 
this  extended detailed balance equation for $P_{eq}(k)$  by
substituting Eqs. \ref{pa11} and \ref{pe11} for $p_a$ and $p_e$ respectively 
in Eqs. \ref{Pa1} and  \ref{Pe1},  leading to the
recurrence relation,
\be
\label{balance2}
P_{eq}(k+1)= \frac{(n-k)p_e(k)}{(k+1)p_a(k+1)}P_{eq}(k)=\frac{B'_e(n-k)(g+n_p)}{B'_a(k+1)(n_p+1)}P_{eq}(k).
\en
This relation can be readily  solved for the equilibrium
distribution $P_{eq}(k)$,  and we obtain 
\be
\label{recurr1}
P_{eq}(k)=(\frac{B'_e}{B'_a})^k\Omega_a(k) \Omega_p(k)\chi,
\en
where
\be
\label{omegaa}
\Omega_a(k) =\frac{n!}{n_g!n_e!}
\en
and
\be
\label{omegap}
\Omega_p(k)=\frac{(g+n_p-1)!}{(g-1)!n_p!},
\en
where $n_g=k$, $n_e=n-k$, and $n_p=n_o+k$, with $n_o=n_q-n$.
We see that $\Omega_a(k)$ is the well-known  expression for the  number of configurations
for the atoms according to Boltzmann's statistics (distinguishable particles), 
while $\Omega_p$ is the corresponding expression for the number of configurations of
photons according to  Bose-Einstein statistics (indistinguishable particles).
The constant   $\chi$ is a normalization factor given by
the condition $\sum P(k)=1$, which yields
\be
\chi ^{-1}= \sum_k (\frac{B'_e}{B'_a})^k\Omega_a(k) \Omega_p(k)
\en
Evidently, to recover Boltzmann's postulate it is also necessary that
$B_a=B_e$. Previously Einstein required this condition 
to derive Planck's black-body formula, but subsequently
it was show to follow directly from quantum electrodynamics.     

If the atoms in this model behave like  fermions, 
we must include the effect of the Pauli exclusion principle
in the probability expressions, Eqs. \ref{Pa1} and \ref{Pe1}.   
Introducing  the  variables  $g_g$ and $g_e$  for the number 
of degenerate ground states and  excited states of the atom,  
we now have
\be
Q^{F}_a(k)= k(g_e-n+k)p_a(k)P^{F}_{eq}(k)\del t,
\en
and
\be
Q^{F}_e(k)=(n-k)(g_g-k)p_e(k)P^{F}_{eq}(k)\del t,
\en
and we obtain  the recurrence relation 
\be
P^{F}_{eq}(k+1)= \frac{B'_e(g_g-k)(n-k)(g+n_p)}{B'_a(g_e-n+k+1)(k+1)(n_p+1)}P^{F}_{eq}(k).
\en
This relation implies that 
\be
\Om^F_a(k)=\frac {g_g!}{(g_g-k)!k!}\frac{g_e!}{(g_e-n+k)!(n-k)!},
\en
which is the  expression for the number of configurations which have
$k$ atoms in the ground state and $n-k$ atoms in the excited state in the case of
Fermi-Dirac statistics \cite{tolman1}.

\subsection*{ III. Computation of the most probable configuration in thermal equilibrium}

At equilibrium, the most probable configuration of a system 
occurs at the maximum  value of $P_{eq}(k)$, Eq. \ref {balance2}.
For large values of $n$,$n_e$,$n_p$ 
and $g$, the Stirling  approximation for the factorial in Eqs. \ref{omegaa} and \ref{omegap}
yields the relations

\be
ln(\Omega_a(k)) \approx -n \, [\frac{n_g}{n}ln(\frac{n_g}{n})+\frac{n_e}{n}ln(\frac{n_e}{n})]
\en
and
\be
ln(\Omega_p(k)) \approx n_p \, [(1+\frac{g}{n_p})ln(1+\frac{g}{n_p})-\frac{g}{n_p}ln(\frac{g}{n_p})].
\en
Setting  $ln(P_{eq}(k))=ln(\Omega_a(k))+ln(\Omega_p(k)) +ln(\chi)$,
we obtain the maximum value for $P_{eq}(k)$ in this approximation   by 
the condition $dln(P_{eq}(k))/dk=0$  which implies
\be 
\label{equi1}
\frac{dln(\Omega_a(k))}{dk}+\frac{dln(\Omega_p(k))}{dk}=0,
\en
where
\be 
\frac{dln(\Omega_a(k))}{dk}=-ln(\frac{n_g}{n_e})
\en
and 
\be
\frac{dln(\Omega_p(k))}{dk}= ln(\frac{g}{n_p}+1).
\en
This condition leads to the  relation   
\be
\label{equ1}
\frac{\bar n_g}{\bar n_e}=\frac{g}{\bar n_p}+1,
\en
where $\bar n_g=k_m$, $\bar n_e=n-k_m$,  and $\bar n_p=n_o+k_m$,
which gives  a quadratic equation for
the most probable value $k_m$ of $k$:
\be
\label{kprob}
k_m=\frac{1}{4}[(n-2n_o-g \pm\sqrt{(n-2n_o-g)^2+8n(g+n_o)} ].
\en
The appropriate  sign for the square root in this solution is determined by the condition that
the most probable number $n_g=k_m$ of atoms in the ground state, 
and the most probable  number of photons $n_p=n_o+k_m$ 
must both be positive. Since the right hand side of Eq. \ref {equ1} is greater than 
one, it is convenient to parameterize this solution by setting
$\bar n_g/\bar n_e= g/\bar n_p+1 =e^{\De}$
where $\De $ is a positive number. Moreover, in accordance with
statistical mechanics, we recognize that we can set 
$\De=\epsilon/k_B T$, where $T$ can be identified 
as the equilibrium temperature of the cavity
and $k_B$ as the  Boltzmann constant. 
Notice  that in this dynamical treatment, $T$ is a positive parameter  
which depends  on the
value of the constants  $n$,$n_o$ and $g$,  
and we recover from first principles Einstein relations, 
Eqs. \ref{statmech}  and \ref{einstein3}, in the form
\be
\frac{\bar n_g}{\bar n_c}=e^{\eps/k_B T}
\en
and 
\be
\bar n_p=\frac{g}{(e^{h\nu/kT}-1)}
\en
for the most probable values of the random
variables $n_g$,$n_e$ and $n_p$.

At this point, it is tempting to identify the ratios $k_m/n=1/Z_a$ and
$(n-k_m)/n=e^{-\epsilon/k_B T}/Z_a$, where $Z_a=1+e^{-\epsilon/k_B T}$, 
with the the canonical  probabilities introduced by Gibbs \cite {gibbs} 
atoms in equilibrium with a thermal bath 
to be in the ground state and  excited  
states, respectively,  where $Z_a$ is the
partition function for the atoms.   
But this is not quite correct, because in our extension of 
Einstein's model, the number of photons  and the number
of atoms in a given configuration  are strictly  
correlated,
while such a correlation is absent if we treat the atoms in equilibrium 
with an external heat bath, as was done originally by Einstein \cite{einst1}.
According to Gibbs, the probability for a configuration that has $k$ atoms
in the ground state and $n-k$ atoms in an excited stated 
in thermal equilibrium with a heat bath at temperature $T$ is given by
\be
\label {gibbs1}
P_G(k)=\Om_a(k) (p_g)^k (p_u)^{n-k}.
\en
Then  the mean value of $k$ is given by
\be
<k>_G= \sum k P_G(k)= \frac{n}{(1+e^{-\eps/k_BT)}}, 
\en
which is equal to $k_m$, and the magnitude of the 
fluctuations of $k$ is given by 
\be
\label{k2G}
<\De k^2>_G=\sum (k^2-<k>^2_G)P_G(k)= n\frac{e^{-\epsilon/k_BT}}
{(1+e^{-\epsilon/k_BT})^2}.
\en
But this result differs from the fluctuations of $k$  obtained by
approximating $P_{eq}(k)$ with  a Gaussian distribution about the mean value
$k_m$. In this latter  case,  one finds that
\be
\label{gausswidth}
\frac{1}{<\De k^2>}=- \frac{d^2ln(\Om_a \Om_p)}{d^2 k}= \frac{n}{n_g n_e}+\frac{g}{n_p(g+n_p)},
\en
evaluated at $k_m=n/(1+e^{-\epsilon/k_BT})$, which  gives
\be
\frac{n_g n_e}{n}= n \frac{e^{-\epsilon/k_BT}}{(1+e^{-\epsilon/k_BT})^2}
\en
and
\be
\frac{n_p(g+n_p)}{g}=\frac{g e^{h\nu/k_BT}}{(e^{h\nu/k_B T}-1)^2}.
\en
The first term is the same  contribution to the 
mean square deviation,\\ $<\De k^2>_G$, Eq. \ref{k2G}, that is
obtained from the  Gibb's probability distribution, Eq. \ref{gibbs1}, 
while the second term is  the corresponding fluctuation $<(\De n_p)^2>_G$ 
that is associated with a gas of photon
in thermal equilibrium. Hence
\be
\frac{1}{<(\De k)^2>}=\frac{1}{<(\De n_g)^2>_G}+\frac{1}{<(\De n_p)^2>_G}.
\en
In Einstein's extended  model, the  fluctuations in the number
of photons can  be neglected provided that the condition
\be
\frac{<(\De n_g)^2>_G}{<(\De n_p)^2>_G}=\frac{(e^{\epsilon/k_BT}-1)^2}
{(e^{\epsilon/k_BT}+1)^2} \frac{n}{g} \ll 1,
\en
is satisfied.  For this inequality to be satisfied at all temperatures, it is 
necessary and sufficient that $n<<g$,
in which case the photons  act as a thermal bath
for the atoms. But for the temperature $T$ to be  determined 
only by the state of the  photon gas, 
it is also necessary that in addition  $n<<n_o$.

To show that the equilibrium configuration obtained at 
the maximum value of the 
probability function $P_{eq}(k)$ corresponds to  thermal equilibrium,
Boltzmann associated the maximum value of the logarithm of the number
of configurations $\Om$ with
the thermodynamic entropy. Setting
\be
\label{enta}
S_a=k_Bln(\Omega_a)
\en
and
\label{entb}
\be
S_p=k_Bln(\Omega_p)
\en
for  the {\it statistical}  entropies  of the atoms
and the photons, respectively, 
the condition for the most probable state of the system, Eq. \ref{equi1}, corresponds
to the Second law of thermodynamics  which states  that at equilibrium the total entropy
$S=S_a+S_b$ of an isolated  system is a maximum.
In  this simple model, the  entropies
$S_a$ and $S_p$ can be expressed as  functions of the  energies
$E_a=n_e\epsilon$ and $E_p=n_p h\nu$, respectively, and 
the temperatures $T_a$ and $T_p$  for the atoms and  photons
in Einstein's model are then given by the  relations
\be
\frac{dS_a}{dE_a}= \frac{1}{T_a}=\frac{k_B}{\epsilon} ln (\frac{n_g}{n_e}),
\en
\be
\frac{dS_p}{dE_p}=\frac{1}{T_p}=\frac{k_B}{h \nu}ln (\frac{g}{n_p}+1).
\en
Thus the maximum condition for the total statistical entropy 
leads to the thermodynamic condition for thermal equilibrium:  
\be
T_a=T_p=T.
\en
We remark that, according to this  definition of temperature, the value 
$T_a$ for the atoms is not restricted to positive values, because the entropy
of the atoms is not a monotonically increasing function of the energy $E_a$.
For example,  $T_a$ becomes  negative when $n_g \le n_e$, but, as we have shown 
above, such a condition
is not possible for atoms in thermal equilibrium with electromagnetic
radiation. Thus, under appropriate conditions the photons can act as 
a thermal bath for the  atoms, but the  atoms cannot provide  a thermal bath
for the  photons.

From the dynamical  point of view
developed here, 
the statistical  entropies $S_a$ and $S_p$,
and the energies $E_a$ and $E_p$ are stochastic variables 
which vary  as a function of time,
as  is the  case also  for the  total entropy $S=S_a+S_b$.   
This is illustrated  in Figs. 1 and 2 which show  
the variation and  fluctuations in the energy and
entropy   as calculated with  the absorption
and emission transition probabilities, Eqs. \ref{pa11} an \ref{pe11}, for
the case that $n=g=n_q=100$. In these calculations
we fixed the  unit of time by setting 
$B=1$, and we determined whether  transitions occur during time intervals 
$\del t=.001$ by using a random number generator. Notice  that in addition to
rapid fluctuations in the energy and  entropy of the atoms and  photons 
there is also a slower and  correlated variation associated with an 
exchange of energy and  entropy between atoms and photons which appears to  oscillate 
irregularly about the most probable value. As expected, the fluctuations  in energy
and entropy about the mean are related approximately by the
thermodynamic condition
\be
\del E_a = T \del S_a
\en
and
\be
\del E_p = T \del S_p.
\en
But these conditions are satisfied here only  approximately  
because the total energy $E=E_a+E_p$
is fixed,  while  the total entropy $S=S_a+S_p$ is not. 
Indeed, as shown  in Fig. 2, the total entropy exhibits  fluctuations which are smaller
than the separate fluctuations in the entropies  of the atoms and the photons, 
and  it is bounded from above by the maximum of the  total entropy.  

\begin{figure}
\includegraphics[width=16cm]{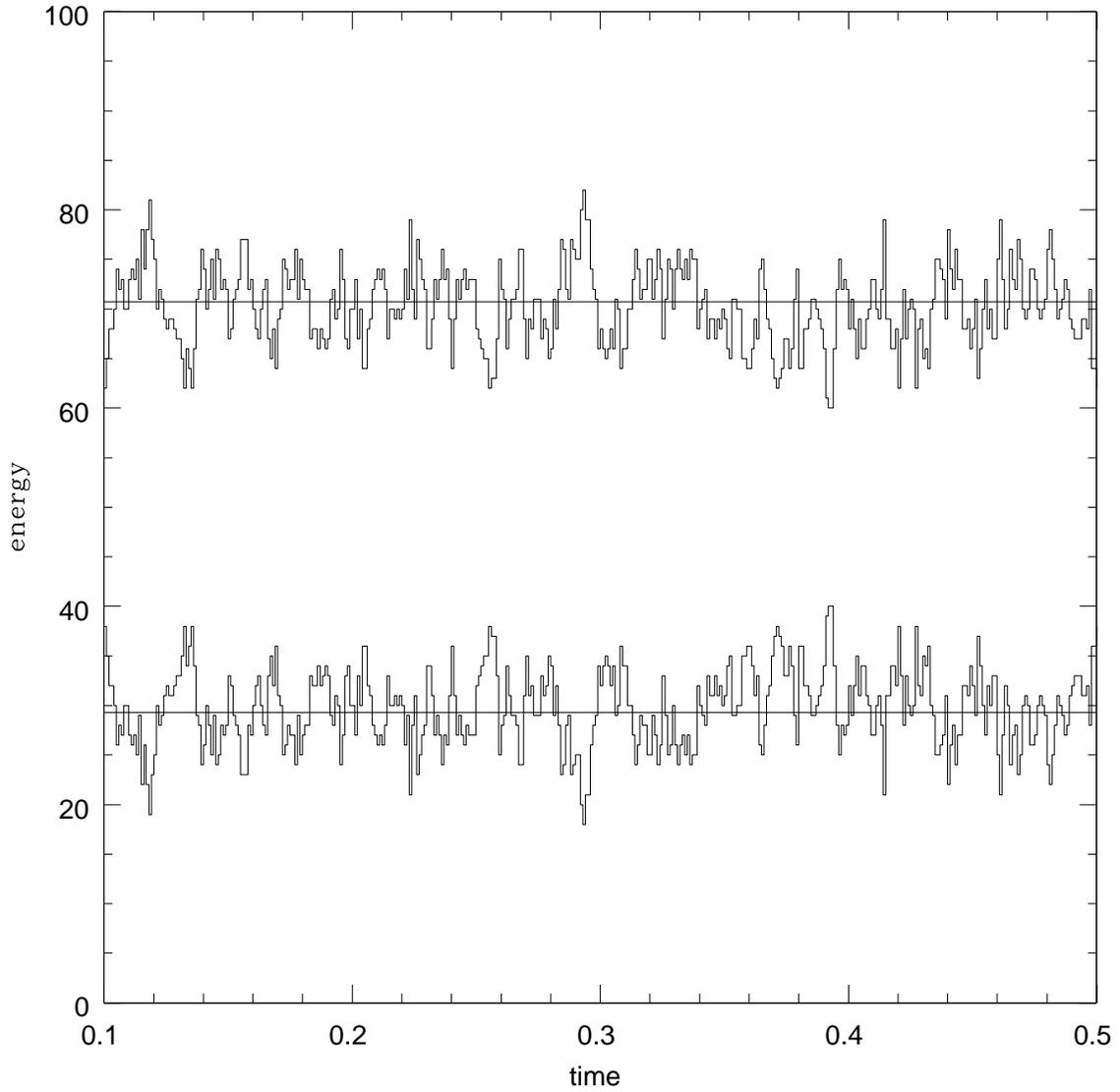}
\caption{\label{} Fluctuations of the energy for atoms (lower curve) and  photons
         (upper curve) in thermal equilibrium}
\end{figure}

\begin{figure}
\includegraphics[width=16cm]{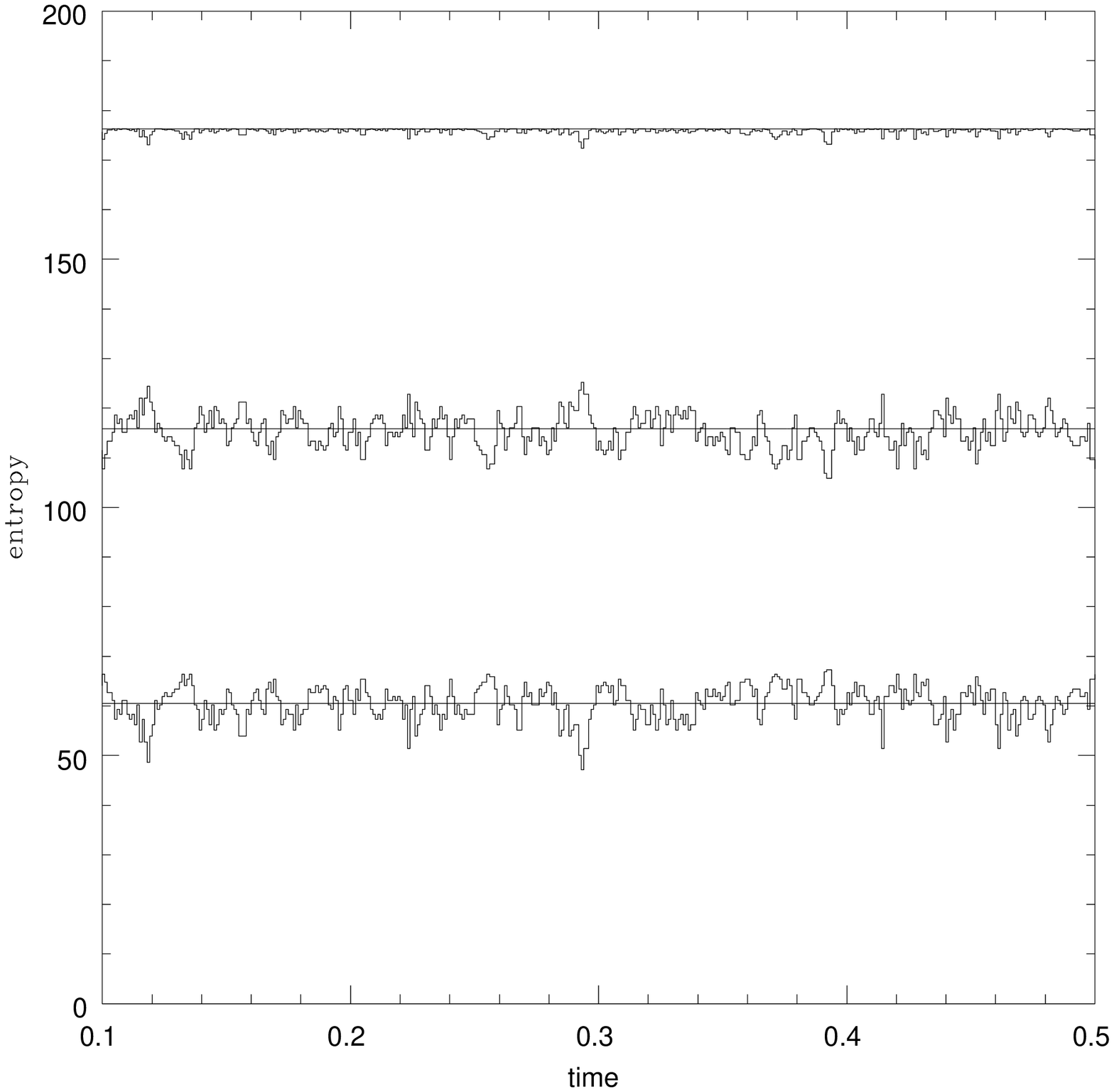}
\caption{\label{} 
 Fluctuations of the entropy for atoms (lower curve) 
 and  photons (middle curve) and for the total entropy (upper curve)  
 in thermal equilibrium}
\end{figure}

\subsection*{IV.  The master equation for the time evolution of 
Einstein's model, and its approach  to thermal equilibrium }

We consider now the time evolution of Einstein's model  
when it is initially in an arbitrary non-equilibrium state. For example,
at $t=0$  all the atoms can be  in the ground state  with
a number of photons in the cavity, or alternatively,  a number of atoms 
can be in the excited state without any photons initially present in the
cavity. Introducing  a  time
dependent probability function $P(k,t)$, Einstein's transition
probabilities for emission and absorption  of radiation  per 
unit time, Eqs. \ref{pa11} and \ref{pe11},   determine  {\it uniquely}  
the probability $P(k,t+\del t)$ at a slightly later  time 
$t+\del t$. We follow here the treatment of stochastic variables 
in van Kampen's book \cite{vankampen}.  
After the small  time interval $\del t$, the configuration
of atoms and  photons represented by the integer $k$ can arrive at $t+\del t$  
under three different conditions: 
\begin{quote}
\item 1) At time $t$ there are $ n_g=k+1$ atoms 
in the ground state, and during the time interval $\del t$ 
one of these atoms absorbs a photon to make a transition to the excited state. 
\item 2) At time $t$ there are $n_e=n-k+1$ atoms in the excited  state, and during the
time interval $\del t$  one of the atoms
in the excited state emits a photon and makes a transition to the ground state.
\item 3) At time $t$  there are $n_g=k$ atoms in the ground state and
$n_e=n-k$ atoms in the excited state, and during the time interval $\del t$  none of these 
atoms absorbs or emits a  photon.
\end{quote}

Adding the probabilities for these three mutually exclusive events, 
we  obtain 
\begin{eqnarray}
\label{master1}
P(k,t+\del t)&=&W(k,k+1)\del t P(k+1,t) + W(k,k-1)\del t P(k-1,t) \nonumber\\
             & & [1 -((W(k+1,k)+W(k-1,k))\del t] P(k,t),
\end{eqnarray}
where 
\be 
W(k-1,k)=k p_a(k) 
\en 
and 
\be
W(k+1,k)=(n-k) p_e(k),
\en
with the boundary  conditions  $W(-1,0)=W(0,-1)=W(n+1,n)=W(n,n+1)=0$.
In the limit $\del t \rightarrow 0$,  one obtains a first order
linear  differential equation for $P(k,t)$ which  can be written in matrix form,
\be
\label{master2}
\frac{dP(k,t)}{dt}=\sum_{k'}W(k,k')P(k',t),
\en
where
\be 
W(k,k)=-k p_a(k)-(n-k) p_e(k).
\en
Note that the matrix elements $W(k',k)$ satisfy the condition $\sum_{k'} W(k',k)=0$, 
as required by the conservation of probability relation $\sum_k dP(k,t)/dt=0$.

The condition $dP_s(k)/dt=0$
for a stationary solution $P_s(k)$ of Eq. \ref{master2} 
is that
\be
\sum_{k'} W(k,k')P_s(k)=0,
\en
which takes the form   
\begin{eqnarray}
& &W(k,k+1) P_s(k+1) -W(k+1,k)P_s(k)=\nonumber\\
& &W(k-1,k) P_s(k)- W(k,k-1)P_s(k-1)=C
\end{eqnarray}
for $0<k<n$, where  
$C$ is a constant independent of $k$, and  for $k=0$ and $k=n$ 
we find that $C=0$. Hence,  
we recover  the extended  detailed balance equation for
the equilibrium distribution $P_{eq}(k)$, Eq. \ref{balance2}, 
and we have $P_s(k)=P_{eq}(k)$

More generally, the solution of the master equation, Eq. \ref{master2}, can be expanded in the form \cite{vankampen}
\be
P(k,t)=P_{eq}(k)+\sum_j c_j(k) e^{-\la_j t}
\en
where the coefficients $c_j(k)$ are eigenvectors of the matrix $W(k,k')$
with eigenvalues $-\la_j$, 
\be
\sum_{k'} W(k,k') c_j(k')=-\la_j c_j(k)
\en
The stationary or equilibrium solution $P_s(k)$  is
a unique  eigenstate of $W(k,k')$ with the eigenvalue $\lambda_0 = 0$.
It can be shown that the other  eigenvalues  $\lambda_j$ are  positive definite  
\cite{vankampen}, 
and therefore all the  solutions of Eq. \ref{master2}  converge to
the equilibrium solution. Another proof for this convergence is given 
in Appendix A by constructing an entropy function which increases
monotonically with time. The length of the eigenvectors  $c_j$ is determined by the initial
conditions $P(k,0)$.  

We illustrate the time evolution of the  probability function $P(k,t)$ 
toward the equilibrium probability  $P_{eq}(k)$, 
by numerically evaluating the solution of the master equation,  Eq. \ref{master2},
for two different initial conditions. In Fig. 3 we consider the
case when initially  there are  $n=200$ atoms in the excited state 
and no photons, so that  $n_q=200$  with $g=200$, and show the probability
function  $P(k,t)$ at the end of each of  10 consecutive time
intervals $\del t=.02$.  
At time  $t=.2$ the solution has nearly approached the equilibrium
solution which according to Eq. \ref{kprob}
has its  maximum at $k_m=100\sqrt{2}$. 
This numerical solution suggests that a very good approximation to $P(k,t)$
is a Gaussian function with time dependent parameters for the mean value
of $k$ and the mean square width $(\De k) ^2$. This is indeed the case, as
will be shown in the next section.
In Fig. 4  we show  the corresponding evolution starting at $t=0$ with
a uniform distribution $P(k,0)=1/n$, with  the same  total energy $n_q=200$, as
in Fig. 3. 
In this case the initial evolution is not represented by a Gaussian function, but 
this  form is seen to be  approached again near the equilibrium. Both cases
evolve to the same final form because we have chosen initially the same total
energy.

\begin{figure}
\includegraphics[width=16cm]{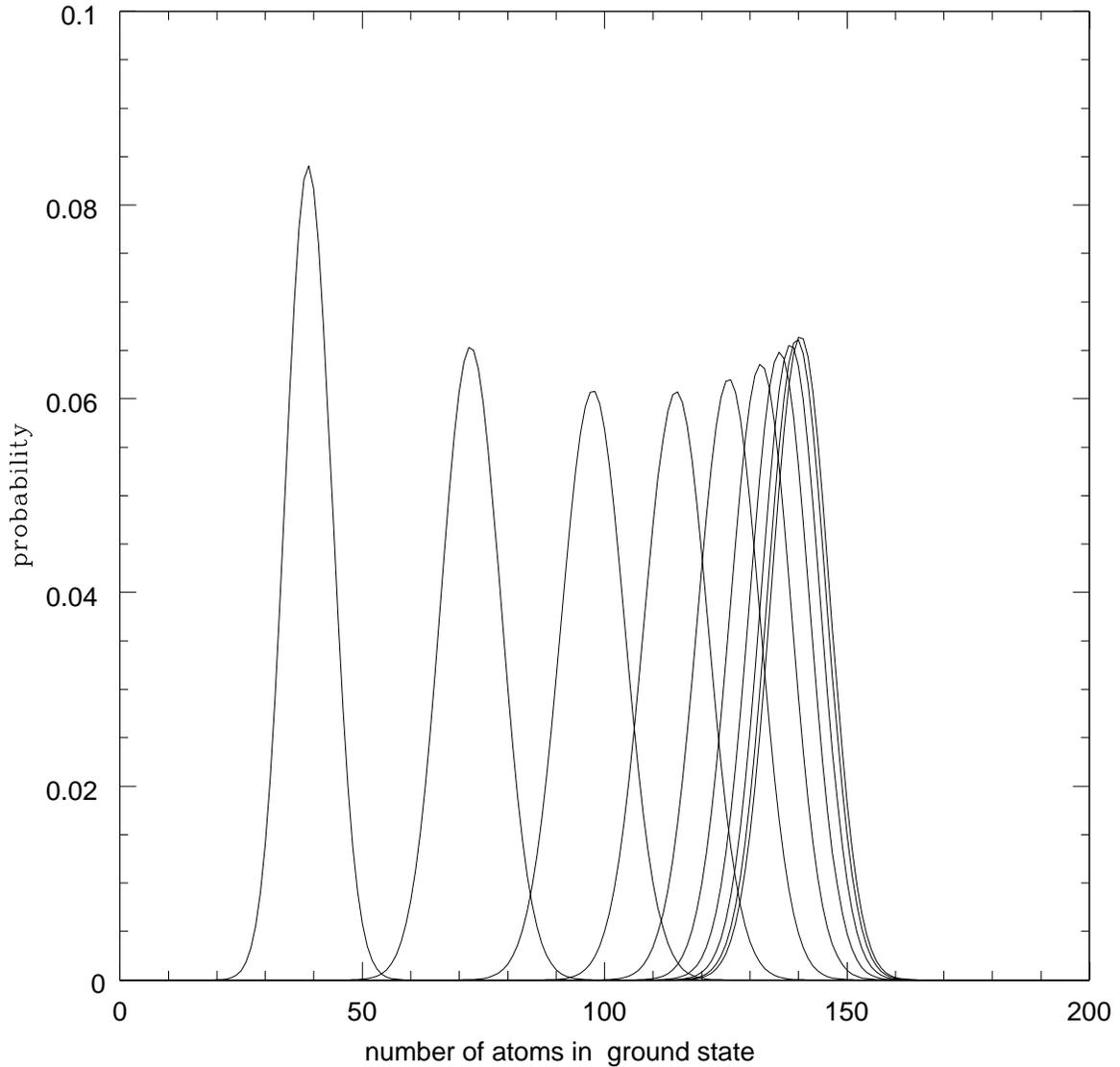}
\caption{
Time evolution of the probability function $P(k,t)$
        at the end of each of  10 consecutive time
        intervals $\del t=.02$, for $g=200$,  when initially 
	there are $n=200$ atoms in the excite state 
        and no  photons.
}
\end{figure}

\begin{figure}
\includegraphics[width=16cm]{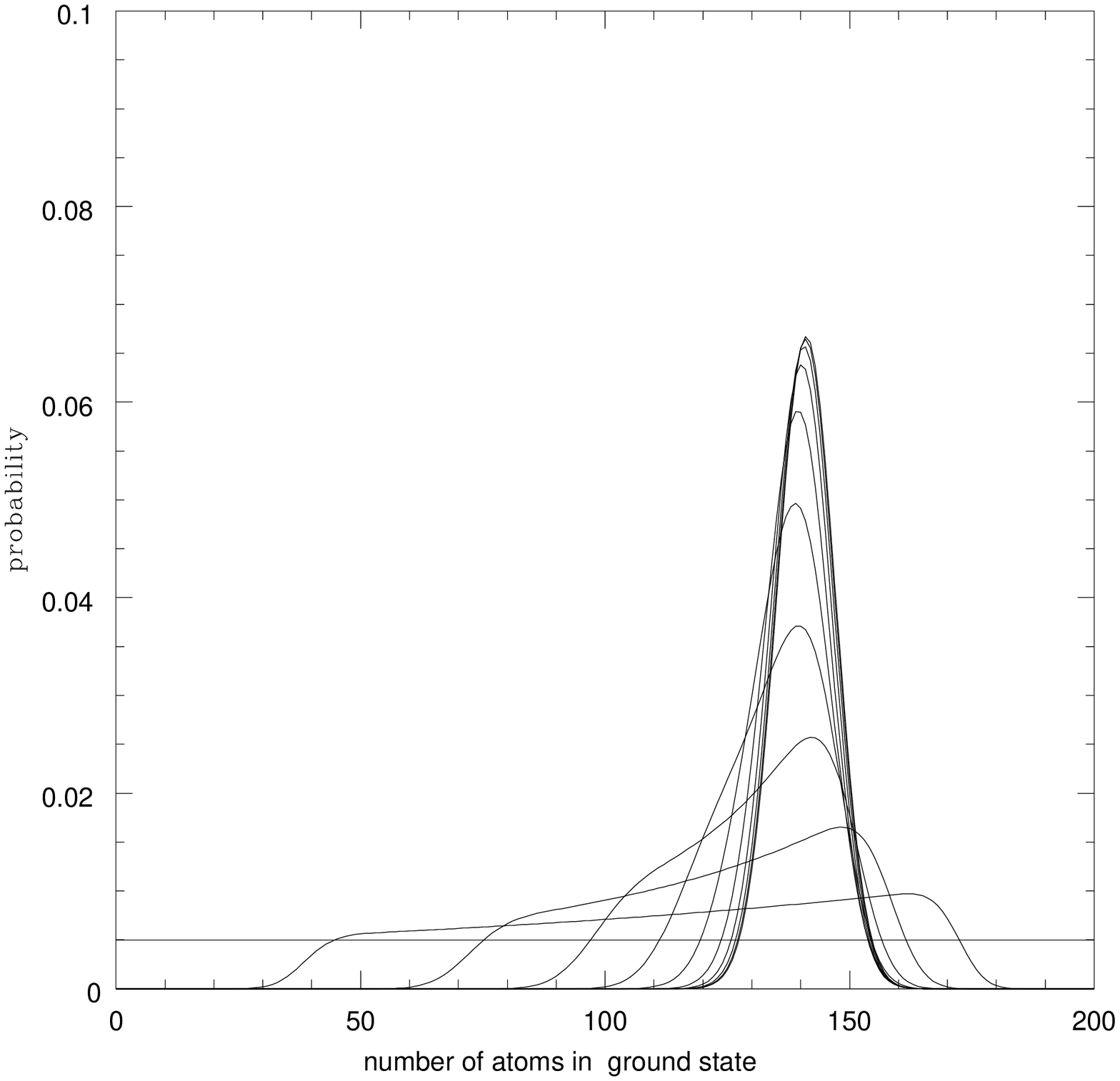}
\caption{
Time evolution for the probability function $P(k,t)$
        at the end of each of  10 consecutive time
        intervals $\del t=.02$, for $g$ and $n=200$, 
	when initially $P(k,0)=1/(n+1)$ and $n_q=200$.
}
\end{figure}

\subsection*{V. Fokker-Planck equation}

Numerical solutions of the master equation, Eq. \ref{master2}, (see for example Fig.3) 
suggests that for large values of the parameters $n$, $n_o$ and $g$,  
the probability  $P(k,t)$ is a continuous  function which  is  well approximated 
by a Gaussian function with  time dependent parameters for the mean value of $k$
and the mean square root  width $\De k$.
We obtain such an approximate solution by
assuming that $P(k,t)$ is a differentiable function
of a continuous variable $k$,  and expanding  $ P(k\pm \del k  ,t)$ to second order
in $\del k$ with  $\del k=1$. Setting
\be
P(k\pm 1,t) \approx P(k,t) \pm \frac{dP(k,t)}{dk}+\frac{1}{2}\frac{d^2 P(k,t)}{dk^2},
\en
we approximate the master equation, Eq. \ref{master2}, by the Fokker-Planck equation
\be
\label{fokpla}
\frac{\partial P(k,t)}{\partial t}=\frac{\partial}{\partial k} [a(k)P(k,t]
+\frac{\partial^2}{\partial k^2}[ b(k)P(k,t)]
\en
where
\be
a(k)=k(k+n_o)-(n-k)(g+n_o+k)
\en
and
\be
b(k)=\frac{1}{2}[k(k+n_o)+(n-k)(g+n_o+k)],
\en
(we have set $B'=1$).
For large values of $n$ and $k$,   this equation can be solved 
approximately by assuming that $P(k,t)$ is a Gaussian function,
\be
P(k,t)=\sqrt{\frac{1}{2\pi  \del (t)}}e^{-(k-k_m(t))^2/2\del (t)}
\en
where  $k_m(t)$ and $\del (t)$ are time dependent parameters.
Then
\be
\frac{\partial P(k,t)}{\partial t}=[-\frac{1}{2\del}\frac{d\del }{dt}
+\frac{(k-k_m)}{\del}\frac{dk_m}{dt}
+\frac{(k-k_m)^2}{ 2 \del^2}\frac{d\del }{dt}]P(k,t),
\en
\be
\frac{\partial P(k,t)}{\partial k}=-\frac{(k-k_m)}{\del}P(k,t),
\en
\be
\frac{\partial ^2P(k,t)}{\partial k^2}=[-\frac{1}{\del}+\frac{ (k-k_m)^2}{\del^2}]P(k,t).
\en
Substituting this expression in the Fokker-Planck equation, 
Eq. \ref{fokpla}, and equating the coefficients of 
$(k-k_m)^j$ for $j=0,1 and2$, we obtain
first order nonlinear differential equations for $k_m(t)$ and $\del (t)$.
Neglecting a term $db(k)/dk$  which is of order $1/n$, we obtain for  
the  $j=1$ terms, 

\be
\label{kder1}
\frac{dk_m(t)}{dt}=-a(k_m(t)),
\en
while for the  $j=0$ and $j=2$ terms  we obtain the {\it same } equation
-a consistency requirement for the validity of  our Gaussian ansatz-, 
\be
\label{derdel}
\frac{d\del(t)}{dt}+2 a'(k_m(t))\del(t)=2 b(k_m(t)),
\en
where
\be
a'(k)=\frac{da(k)}{dk}=4k+2n_o-g-n.
\en
These two equations can also be obtained  from the master equation, Eq. \ref{master2}, by 
evaluating the time derivatives of the averages $k_m(t)=<k>$ and 
$\del(t)=<(\De k)^2>$, 
assuming that the probability function $P(k,t)$ is sharply peaked at $k=k_m(t)$. 

Assuming that initially we have $k=k_m(0)$, where  $0\leq k_m(0) \leq n$,  
the solution of Eq. \ref{kder1} is 
given by 
\be
\label{kmt}
k_m(t)=\frac{(k_{+}-k_{-}\phi\, e^{-\lambda t})}{(1-\phi\, e^{-\lambda t})},
\en
where
\be
\phi=\frac{(k_m(0)-k_{+})}{(k_m(0)-k_{-})},
\en
\be
\lambda=\sqrt{(n-2n_o-g)^2+8n(n_o+g)},
\en
and
\be
k_{\pm}=\frac{1}{4}(n-2n_o-g \pm \lambda).
\en
Here time is measured in units of $1/B'$, and in the limit
$\la t >>1$ we see that $k_m$ approaches $k_{+}$, which is equal to
the most probable value of $k$ at equilibrium obtained previous, Eq. \ref{kprob}. 

The solution for $\del (t)$, Eq. \ref{derdel}, where initially
$\del(0)=\del_o$, is given by 
\be
\label{delt}
\del (t)= \frac{1}{n}e^{-\xi(t)}\int_0^t dt' e^{\xi(t')}b(k_m(t'))+\del_o e^{-\xi(t)}
\en
where
\be
\xi(t)=\frac{2}{n} \int_0^t dt'a'(k_m(t')). 
\en
For $\lambda t>>1 $, we have  $\xi(t) \approx 2 \lambda t$ and
$\del (t)$ approaches the equilibrium  value 
\be
\del = \frac{b(k_{+})}{a'(k_{+})}=\frac{k_{+}(n-g)+n(g+n_o)}{2(4k_{+}+2n_o+g-n)},
\en
which can be shown to correspond to the 
value for $<(\De k)^2>$ at equilibrium,  Eq. \ref{gausswidth}
obtained previously.

In the  limit that both $n_o$ and $g$ are  much  larger than $n$,
the  changes in  the numbers of photons in the cavity
can be neglected in the expression for the probabilities for the absorption and
emission of photons, Eqs. \ref{pa11} and \ref{pe11}, and  we have 
\be
p_a \approx B' n_o
\en
and
\be
p_e \approx B'(g+n_o)
\en
In this case
the photons act as a { \it heat bath} for the atoms
at  a temperature  determined by $g$ and $n_o$,  where
$T=(1/h\nu)ln(g/n_o+1)$.  
In this limit, the time dependent solution, Eqs. \ref{kmt} 
and \ref{delt},  simplifies to the form 
\be 
\label{kmt2}
k_m(t)=np+(k_m(0)-np)e^{-\lambda t} 
\en 
and 
\be 
\label{delt2}
\del(t)=\frac{n_o}{\la}[p(1-e^{-2\la t})+(1-e^{\De})(\frac{k_m(0)}{n} -p)e^{-\la t}(1-e^{-\la t})], 
\en 
where $\la=n_o(e^{h\nu/k_B T}+1)$ and $p=1/(1+e^{-h\nu/k_B T})$ 
is  the canonical probability of finding an atom in the ground state
at temperature $T$.

For systems in contact with a thermal bath, our model  
explains the puzzle  that a ratio of  transition 
probabilities like  $p_e/p_a$,which is determined by the 
underlying dynamics of the system, can also be expressed in terms of the  
canonical Gibbs probability function, which depends  on the temperature
of the heat bath \cite {kamp2}.  Indeed,  as we have shown here, the temperature of the 
heat bath is determined by the magnitude of this ratio. In this limit we find that Enstein's model, 
in the version discussed here,  corresponds mathematically
to the Ehrenfests `dog-flea'' model \cite{ehren}, which was  extended recently  
to finite temperatures by Ambegaokar and Clerk \cite{ambegaokar}. 
Hence Eqs. \ref{kmt2} and \ref{delt2} provide also an approximate analytic solution 
for this model, which previously had been solved by M. Kac \cite{kac} 
only for the special case that  $p=1/2$ or infinite temperature.

\subsection*{VI. Summary } 

We have shown that a stochastic treatment of a simplified  version of Enstein's 1916  
model for atoms in interaction with photons, which  conforms with modern quantum 
electrodynamics, serves to  illustrate various  aspects of the foundations of statistical 
mechanics. From the underlying quantum dynamics of this model  
we obtained directly  the equilibrium probability distribution $P_{eq}$ 
in accordance with the  familiar results obtained 
in statistical mechanics.
For atoms treated as  distinguishable entities, we found that  
this equilibrium  probability is  proportional to 
the number of configurations associated with Boltzmann's statistics,
while  for  atoms which obey the Pauli exclusion principle, 
the  corresponding statistics is   Fermi-Dirac,  and
for  photons it is  Bose-Einstein.
Conventionally, these probability  distributions are derived 
in statistical mechanics by using  a {\it postulate} introduced by
Boltzmann \cite {boltz1} in 1877, which states that at equilibrium all the  microstates of a system
at a fixed energy -  a micro canonical ensemble in the language of Gibbs \cite {gibbs}-
are equally probable. 

For  the time evolution of 
the Einstein  model we discussed  a master equation  which describes the approach
of the probability distribution 
to a {\it unique} stationary solution
which corresponds to the  equilibrium distribution,
starting  from any arbitrary initial state. 
In the  Fokker-Planck approximation  we obtained an analytic
solution of this equation which is in excellent  agreement with 
the numerical solutions some of which are presented here.
Associated with this probability function we described  an 
entropy function  
which increases monotonically in time reaching a maximum value 
at the  thermal equilibrium distribution. 

We must add, however, an important caveat. In quantum
mechanics a state of an isolated system can be  described 
by a wavefunction, and at any time $t$ when there are  atoms in both the ground state and  
the excited state, as well as  photon states, this wave function is a linear
combination of wave functions describing these states. 
Therefore transition probabilities, which are  
obtained from bilinear terms of this wavefunction, contain  interference terms
that have been  ignored  here.  In the literature this is known as the random phase 
approximation or decoherence,
but there is no consensus as how to justify this approximation from first principles,
and a deeper understanding is lacking. Moreover, there is the famous problem with 
the  measurement process associated with the Copenhagen interpretation 
of quantum mechanics.  
Such a process which is indicated by the application  
of transition probabilities  repeated here  {\it ad infinitum}  
during short intervals of time $\del t$, is obviously not  
taking place in a macroscopic system that is evolving towards  thermodynamic equilibrium,
unless one wants to believe that there is a Maxwell-like demon   
who is continuously carrying 
out such measurements. In our view, however, 
the  success of the detailed balance approach described here
justifies the application of the Copenhagen interpretation to transition
probabilities per unit time also in  the {\it  absence}  of any identifiable
measurement process and/or the existence of an {\it observer}. 

From a historical perspective , it
is interesting to speculate  that Einstein, who was the master
of fluctuation theory, could also  have carried out  calculations
similar to the ones  we have done here, thus  resolving early-on  
the bitter conflicts which exist(ed) 
concerning the statistical approach to thermal equilibrium proposed by Boltzmann,
for systems obeying time-symmetric dynamics.
Recently, Ambegaokar and Clerk \cite{ambegaokar} 
discussed the  1906 ``dog-flea'' model 
the Ehrenfests \cite{ehren}, which  provided an early
model to illustrate Boltzmann's ideas.
In a special limit, we have shown  that the extension of this model to finite temperatures  \cite{ambegaokar}  
is mathematically equivalent to  Einstein's model  for atoms interacting with radiation, 
when the atoms have only two levels.
An analytic solution for the  original  Ehrenfests' model
was not obtained  until 1947 by Kac \cite {kac}, 
and we have now also  obtained an analytic  solution, 
in the Fokker-Planck approximation, Eqs. \ref{kmt2} and \ref{delt2},
for the finite temperature extension of this model \cite{ehren1}.

\subsection*{Appendix A. Entropy}
We  discuss  the definition of  entropy and its evolution in time 
when Einstein's  model is not initially in equilibrium. It
would appear that a natural extension 
for entropy is the quantity $\sum_k P(k,t)ln\Om(k)$, but this expression does not 
have the desired  property that the entropy  increase monotonically with time. Instead, 
this property is satisfied by a related expression \cite{vankampen} \cite{voneuman},
\be
\label{entropy2}
S(t)=-\sum_k P(k,t)ln\frac{P(k,t)}{\Om(k)}.
\en
In the case that  $P(k,t)$ is sharply peaked at some value $k(t)$,   
as we have found  when $P(k,t)$ approaches the equilibrium
distribution $P_{eq}(k)$, the term 
\be
-\sum_k P(k,t)ln P(k,t)
\en
is small compared to 
\be
\sum_k P(k,t)ln \Om(k).
\en
In turn, this term can then be approximated by $ln \Om(k(t))$,
and we have for  the entropy $S(t)$, 
\be
S(t) \approx ln\Om(k(t)).
\en

Applying the master equation, Eq. \ref{master2}, to the definition for
entropy, Eq. \ref{entropy2}, we obtain
\be
\frac{dS(t)}{dt}=-\sum_{k,k'}W(k,k')P(k',t) ln\frac{P(k,t)}{\Om(k)}.
\en
Substituting the relation
\be
W(k,k)=-(W(k+1,k)-W(k-1,k),
\en
and setting
\be
Z(k,t)=\frac{P(k+1,t)}{\Om(k+1)}-\frac{P(k,t)}{\Om(k)},
\en
we obtain
\be
\frac{dS(t)}{dt}=\sum_k[W(k+1,k)P(k,t)-W(k,k+1)P(k+1,t)]ln Z(k,t).
\en
Substituting the relation
\be
W(k,K+1)\Om(k+1)=W(k+1,k)\Om(k),
\en
we obtain
\be
\frac{dS(t)}{dt}=\sum_k W(k+1,k)\Om(k)Z(k,t)lnZ(k,t),
\en
which proves that \cite{vankampen}  
\be
\frac{dS(t)}{dt} \geq 0.
\en

Thus, the entropy  defined in Eq.\ref{entropy2} 
increases monotonically towards a maximum value corresponding
to  thermal equilibrium.
Recently proposed  generalizations of this form of the
entropy \cite{michael2} are incompatible with this fundamental requirement.

\subsection*{Acknowledgements}

I would like to thank Vinay Ambegaokar for his insightful comments, and
for  calling my attention to the discussion of the Ehrenfests model 
by M. Kac \cite{kac}.


\begin{thebibliography}{30}
\bibitem{ehren1}      
To demonstrate  the monotonic increase of total entropy
in the finite temperature  extension of the
Ehrenfests model one must
appeal to the relation between
entropy and heat change for systems in thermal equilibrium
\cite{ambegaokar}.
But this is unnecessary in our treatment of the Einstein
model where  the heat bath is
an intrinsic part of a dynamical system.
\end{thebibliography}
\end{document}